\documentclass[twocolumn,prc,10pt,amsmath,amssymb,aps,showpacs,superscriptaddress]{revtex4-1}

\usepackage{graphicx}
\usepackage{dcolumn}
\usepackage{bm}
\usepackage{}[multicol]
\usepackage{color}
\newcommand{\eek}{$(e,e^{\prime}K^{+})$}
\newcommand{\pik}{$(\pi^{+},K^{+})$}
\newcommand{\kpi}{$(K^{-},\pi^{-})$}
\newcommand{\cerenkov}{\v{C}erenkov}
\newcommand*{\TOHOKU}{Department of Physics, Graduate School of Science, Tohoku University, Sendai, Miyagi 980-8578, Japan}
\newcommand*{\MAINZ}{Institute for Nuclear Physics, Johannes Gutenberg-University, D-55099 Mainz, Germany}
\newcommand*{\NCarolina}{Department of Physics, North Carolina A$\&$T State University,Greensboro, NC 27411, USA}
\newcommand*{\Hampton}{Department of Physics, Hampton University, Hampton, VA 23668, USA}
\newcommand*{\Zagreb}{Department of Physics $\&$ Department of Applied Physics, University of Zagreb, HR-10000 Zagreb, Croatia}
\newcommand*{\Yervan}{A.I.Alikhanyan National Science Laboratory, Yerevan 0036, Armenia}
\newcommand*{\FIU}{Department of Physics, Florida International University, Miami, FL 27411, USA}
\newcommand*{\Christpher}{Department of Physics, Computer Science $\&$ Engineering, Christopher Newport University, Newport News, VA, USA 23606}
\newcommand*{\JLAB}{Thomas Jefferson National Accelerator Facility (JLab), Newport News, VA 23606, USA}
\newcommand*{\Bari}{Istituto Nazionale di Fisica Nucleare, Sezione di Bari and University of Bari, I-70126 Bari, Italy}
\newcommand*{\Southern}{Department of Physics, Southern University at New Orleans,New Orleans, LA 70126, USA}
\newcommand*{\NCarolinatwo}{Department of Physics, University of North Carolina at Wilmington, Wilmington, NC 28403, USA}
\newcommand*{\Roma}{INFN, Sezione Sanit$\grave{a}$ and Istituto Superiore di Sanit$\grave{a}$, 00161 Rome, Italy}
\newcommand*{\Lanzhou}{Nuclear Physics Institute, Lanzhou University, Gansu 730000, China}
\newcommand*{\Mississippi}{Department of Physics, Mississippi State University, Mississippi State, Mississippi 39762, USA}
\newcommand*{\James}{Department of Physics and Astronomy, James Madison University, Harrisonburg, VA 22807, USA}
\newcommand*{\Rico}{Escuela de Ciencias y Tecnologia, Universidad Metropolitana, San Juan, 00928, Puerto Rico}
\newcommand*{\VMI}{Department of Physics $\&$ Astronomy, Virginia Military Institute, Lexington, Virginia 24450, USA}
\newcommand*{\Yamagata}{Department of Physics, Yamagata University,Yamagata, 990-8560, Japan}
\newcommand*{\Xavier}{Department of Physics, Xavier University of Louisiana,New Orleans, LA 70125, USA}
\newcommand*{\Hou}{Department of Physics, University of Houston, Houston, Texas 77204, USA}

\begin{document}
\title{High Resolution Spectroscopic Study of $^{10}_{\Lambda}$Be}

\author{T.~Gogami}
\thanks{{\it Current address: Department of Physics, Kyoto University, Kyoto, 606-8502, Japan}}
\affiliation{\TOHOKU}
\author{C.~Chen}
\affiliation{\Hampton}
\author{D.~Kawama}
\affiliation{\TOHOKU}
\author{P.~Achenbach}
\affiliation{\MAINZ}
\author{A.~Ahmidouch}
\affiliation{\NCarolina}
\author{I.~Albayrak}
\affiliation{\Hampton}
\author{D.~Androic} 
\affiliation{\Zagreb}
\author{A.~Asaturyan}
\affiliation{\Yervan}
\author{R.~Asaturyan}\thanks{Deceased}
\affiliation{\Yervan}
\author{O.~Ates} 
\affiliation{\Hampton}
\author{P.~Baturin}
\affiliation{\FIU}
\author{R.~Badui}
\affiliation{\FIU}
\author{W.~Boeglin}
\affiliation{\FIU}
\author{J.~Bono}
\affiliation{\FIU}
\author{E.~Brash} 
\affiliation{\Christpher}
\author{P.~Carter}
\affiliation{\Christpher}

\author{A.~Chiba}
\affiliation{\TOHOKU}
\author{E.~Christy}
\affiliation{\Hampton}
\author{S.~Danagoulian}
\affiliation{\NCarolina}
\author{R.~De~Leo} 
\affiliation{\Bari}
\author{D.~Doi} 
\affiliation{\TOHOKU}
\author{M.~Elaasar}
\affiliation{\Southern}
\author{R.~Ent} 
\affiliation{\JLAB}
\author{Y.~Fujii}
\affiliation{\TOHOKU}
\author{M.~Fujita}
\affiliation{\TOHOKU}
\author{M.~Furic}
\affiliation{\Zagreb}
\author{M.~Gabrielyan}
\affiliation{\FIU}
\author{L.~Gan}
\affiliation{\NCarolinatwo}
\author{F.~Garibaldi}
\affiliation{\Roma}
\author{D.~Gaskell}
\affiliation{\JLAB}
\author{A.~Gasparian} 
\affiliation{\NCarolina}
\author{Y.~Han}
\affiliation{\Hampton}
\author{O.~Hashimoto}\thanks{Deceased}
\affiliation{\TOHOKU}
\author{T.~Horn}
\affiliation{\JLAB}
\author{B.~Hu}
\affiliation{\Lanzhou}
\author{Ed.V.~Hungerford}
\affiliation{\Hou}
\author{M.~Jones} 
\affiliation{\JLAB}
\author{H.~Kanda}
\affiliation{\TOHOKU}
\author{M.~Kaneta}
\affiliation{\TOHOKU}
\author{S.~Kato}
\affiliation{\Yamagata}
\author{M.~Kawai}
\affiliation{\TOHOKU}

\author{H.~Khanal}
\affiliation{\FIU}
\author{M.~Kohl} 
\affiliation{\Hampton}
\author{A.~Liyanage}
\affiliation{\Hampton}
\author{W.~Luo}
\affiliation{\Lanzhou}
\author{K.~Maeda}
\affiliation{\TOHOKU}
\author{A.~Margaryan} 
\affiliation{\Yervan}
\author{P.~Markowitz}
\affiliation{\FIU}
\author{T.~Maruta}
\affiliation{\TOHOKU}
\author{A.~Matsumura}
\affiliation{\TOHOKU}
\author{V.~Maxwell}
\affiliation{\FIU}
\author{A.~Mkrtchyan}
\affiliation{\Yervan}
\author{H.~Mkrtchyan}
\affiliation{\Yervan}
\author{S.~Nagao}
\affiliation{\TOHOKU}
\author{S.N.~Nakamura}
\affiliation{\TOHOKU}
\author{A.~Narayan} 
\affiliation{\Mississippi}
\author{C.~Neville}
\affiliation{\FIU}
\author{G.~Niculescu} 
\affiliation{\James}
\author{M.I.~Niculescu}
\affiliation{\James}
\author{A.~Nunez}
\affiliation{\FIU}
\author{Nuruzzaman}
\affiliation{\Mississippi}
\author{Y.~Okayasu} 
\affiliation{\TOHOKU}
\author{T.~Petkovic}
\affiliation{\Zagreb}
\author{J.~Pochodzalla}
\affiliation{\MAINZ}
\author{X.~Qiu}
\affiliation{\Lanzhou}
\author{J.~Reinhold}
\affiliation{\FIU}
\author{V.M.~Rodriguez}
\affiliation{\Rico}
\author{C.~Samanta}
\affiliation{\VMI}
\author{B.~Sawatzky} 
\affiliation{\JLAB}
\author{T.~Seva}
\affiliation{\Zagreb}
\author{A.~Shichijo}
\affiliation{\TOHOKU}
\author{V.~Tadevosyan}
\affiliation{\Yervan}
\author{L.~Tang}
\affiliation{\Hampton}
\affiliation{\JLAB}
\author{N.~Taniya}
\affiliation{\TOHOKU}
\author{K.~Tsukada}
\affiliation{\TOHOKU}
\author{M.~Veilleux}
\affiliation{\Christpher}
\author{W.~Vulcan}
\affiliation{\JLAB}
\author{F.R.~Wesselmann}
\affiliation{\Xavier}
\author{S.A.~Wood} 
\affiliation{\JLAB}
\author{T.~Yamamoto}
\affiliation{\TOHOKU}
\author{L.~Ya}
\affiliation{\Hampton}
\author{Z.~Ye}
\affiliation{\Hampton}
\author{K.~Yokota}
\affiliation{\TOHOKU}
\author{L.~Yuan} 
\affiliation{\Hampton}
\author{S.~Zhamkochyan}
\affiliation{\Yervan}
\author{L.~Zhu}
\affiliation{\Hampton}

\collaboration{ HKS(JLab E05-115) Collaboration }


\begin{abstract}
  Spectroscopy of a $^{10}_{\Lambda}$Be hypernucleus 
  was carried out at JLab Hall C using the {\eek} reaction.
  A new magnetic spectrometer system (SPL+HES+HKS), specifically designed
  for high resolution hypernuclear spectroscopy, was used to obtain an 
  energy spectrum with a resolution of $\sim 0.78$~MeV (FWHM).
  The well-calibrated spectrometer system of the present experiment
  using $p${\eek}$\Lambda$,$\Sigma^{0}$ 
  reactions allowed us to determine the energy levels, 
  and the binding energy of 
  the ground state peak (mixture of $1^{-}$ and $2^{-}$ states) was 
  obtained to be $B_{\Lambda} = 8.55 \pm 0.07({\rm stat.}) \pm 0.11({\rm sys.})$~MeV.
  The result indicates that the 
  ground state energy is shallower 
  than that of an emulsion study by about 0.5~MeV
  which provides valuable experimental information on the
  charge symmetry breaking effect in the $\Lambda N$ interaction. 
\end{abstract} 
\maketitle
\section{Introduction}
Knowledge of the Nucleon-Nucleon ($NN$) system can be generalized to 
the Baryon-Baryon ($BB$) system using SU(3) flavor symmetry. 
A study of the Hyperon-Nucleon ($YN$) system is a natural 
extension as a first step from $NN$ to $BB$ system.
However, a $YN$ scattering experiment, which is the most 
straightforward way to explore the interaction, is very limited 
due to the short lifetimes of hyperons ($e.g.$ $\tau=263$~ps for $\Lambda$).
Thus, there has been interest in hyperon binding in a nucleus where
the $YN$ interaction can be studied with higher precision
and greater statistical accuracy.  
The spectroscopy of $\Lambda$ hypernuclei was begun approximately 60
years ago following the serendipitous discovery of a hypernucleus
in an emulsion exposed to cosmic rays~\cite{cite:danysz} 
to investigate the $\Lambda N$ interaction.
Counter experiments initially used either the {\kpi} and
{\pik} reactions in facilities located in CERN, BNL and KEK. 
A number of $\Lambda$ hypernuclei have been observed up
to a nuclear mass of $A=209$, and new features which do not present in the ordinary 
nuclear system were observed in the hypernuclear system~\cite{cite:hashimototamura}.
One of the most novel results from the hypernuclear spectroscopy 
is a clear evidence of the nuclear shell structures even for 
deep orbital states in heavy mass nuclei, 
measured by the {\pik} experiments~\cite{cite:hashimototamura}. 
This is thanks to a fact that a single embedded $\Lambda$ is 
not excluded from occupying inner nuclear shells 
by the Pauli principle from nucleons. 
Thus, the $\Lambda$ can dynamically probe the nuclear interior as an impurity.
However, more accurate and detailed 
structures for a variety of hypernuclei are still
needed to improve our understanding of the strongly-interacting system with 
a strangeness degree of freedom, and 
they are being tried to be 
measured in complementary ways at J-PARC
using hadron beams~\cite{cite:jparc},
GSI (FAIR) using heavy ion beams~\cite{cite:hyphi}, 
MAMI~\cite{cite:patrick} and 
JLab using electron 
beams~\cite{cite:miyoshi,cite:lulin,cite:iodice,cite:cusanno,cite:cusanno2,cite:28LAl,cite:7LHe,cite:12LB}
today.

\section{Missing Mass Spectroscopy of $\Lambda$ Hypernuclei with Electron Beams}
Missing mass spectroscopy using the {\eek} 
reaction started in 2000 at Thomas Jefferson National Accelerator Facility 
(JLab)~\cite{cite:miyoshi,cite:lulin}. 
The missing mass spectroscopy with primary electron beams
enabled us to have better energy resolution in the
resulting hypernuclear structure
than that with currently available hadron beams (FWHM of a few MeV).
These initial studies demonstrated that FWHM~$\simeq0.9$~MeV resolution was possible.
Then, experiments at JLab in 2005 and 2009
(JLab E01-011~\cite{cite:28LAl,cite:7LHe,cite:12LB} and E05-115~\cite{cite:12LB}) were
performed with new magnetic spectrometers specifically designed for 
a resolution of FWHM~$\simeq 0.5$~MeV 
which is the best in the missing mass spectroscopy of the $\Lambda$ hypernuclei.
In terms of the energy resolution, 
the gamma-ray spectroscopy~\cite{cite:hashimototamura}, which measures de-excitation gamma-rays
from hypernuclei, is better (typically FWHM$\simeq$ a few keV) than that of the 
missing mass spectroscopy. 
$\Lambda$ hypernuclei with the light mass numbers ($A\leq19$)
were investigated by the gamma-ray spectroscopy with such a high 
energy resolution up to now. 
However, it measures only energy-level spacing 
and cannot give absolute binding energy and production cross section information. 
Therefore, missing mass measurements and gamma-ray spectroscopy are complementary, 
and both of them are important.

In addition to providing higher resolution in the missing mass spectroscopy,
the {\eek} reaction has other useful features. 
An absolute mass-scale can be calibrated with the 
$p${\eek}$\Lambda$,$\Sigma^{0}$ reactions, since all masses are known~\cite{cite:pdg},
and in particular, the reaction constituents lead to clearly resolved peaks
which are well separated from backgrounds. Furthermore, the {\eek} reaction
converts a nuclear proton into a
$\Lambda$ in contrast to mesonic reactions which convert a neutron into
a $\Lambda$. Comparison of these isotopic mirror hypernuclei 
provides information on
hypernuclear charge symmetry.
This paper reports the first result of a spectroscopic measurement of 
$^{10}_{\Lambda}$Be with the new spectrometers 
that allowed us to observe the finer nuclear structures and 
to determine the energy levels with a better precision 
than those of the previous $\Lambda$ hypernuclear missing mass 
spectroscopy. 

\section{Motivation for the Measurement of $^{{\bf 10}}_{{\bf\Lambda}}{\rm {\bf Be}}$}
The present study compares the ground state binding energies of 
$^{10}_{\Lambda}$Be ($\alpha + \alpha + n + \Lambda$) 
to that of its isotopic mirror nucleus
$^{10}_{\Lambda}$B ($\alpha + \alpha + p + \Lambda$), providing
information on the mechanism of Charge Symmetry Breaking (CSB) in hypernuclei.
The most evident fact of the $\Lambda N$ CSB is the ground state binding energy difference 
of $A=4$ iso-doublet ($T=1/2$) hypernuclei, 
$B_{\Lambda}(^{4}_{\Lambda}{\rm He}) - B_{\Lambda}(^{4}_{\Lambda}{\rm H}) 
= (2.39 \pm 0.03) - (2.04\pm0.04)= +0.35\pm0.06$~MeV~\cite{cite:12LC_2}.
The $\Lambda$ binding energy is defined by using 
masses of $\Lambda$ ($M_{\Lambda}$) and core nucleus, 
for example 
$B_{\Lambda}(^{4}_{\Lambda}{\rm He})=M(^{3}{\rm He})+M_{\Lambda}-M(^{4}_{\Lambda}{\rm He})$. 
The binding energy difference is larger beyond our expectations even 
after a correction in the change in the Coulomb energy due to core contraction, 
and it is attributed to the $\Lambda N$ CSB effect~\cite{cite:bodmer,cite:gibson}.
Many theorists have tried to understand the origin of the $\Lambda N$ CSB
for more than 40 
years~\cite{cite:hiyama_10LB,cite:gibson,cite:bodmer,cite:hiyama1,cite:gal,cite:nogga,cite:akaishi}, but still it has not fully been understood yet.
Recently, the binding energy of $^{4}_{\Lambda}$H was 
remeasured by a decayed pion spectroscopy at MAMI
to confirm the emulsion measurement~\cite{cite:patrick}. 
The result of $B_{\Lambda}(^{4}_{\Lambda}{\rm H}) = 2.12 \pm 0.01 \pm 0.09$~MeV
is consistent with that of the emulsion experiment.
Results from a recent gamma-ray spectroscopy measurement indicated 
that the CSB effect of 1$^{+}$ states in the $A=4$, $T=1/2$ hypernuclear system
is small, which demonstrated the CSB interaction 
strongly depends on the spin~\cite{cite:yamasan}.
The study of $\Lambda N$ CSB effect in $p$-shell hypernuclear system is 
also under the spotlight these days~\cite{cite:7LHe,cite:hiyama_10LB,cite:hiyama1,cite:gal}. 
A study for the $A=7$, $T=1$ $\Lambda$ hypernuclei, 
which are of the simplest $p$-shell system,
show that the experimental results~\cite{cite:7LHe} 
do not favor a phenomenological $\Lambda N$ CSB potential
in cluster model~\cite{cite:hiyama1}. 
The similar study in $A=10$, $T=1/2$ system was
performed by comparing the ground state binding energies of 
$^{10}_{\Lambda}$Be and $^{10}_{\Lambda}$B~\cite{cite:hiyama_10LB,cite:gal}.
However, the reported binding energy of $^{10}_{\Lambda}$Be
was a weighted-mean value of only three events 
measured in the emulsion experiment~\cite{cite:juric,cite:cantwell}
though the binding energy of $^{10}_{\Lambda}$B was determined 
by higher statistic (nevertheless there are only ten events).
Thus, the binding energy measurement of $^{10}_{\Lambda}$Be with 
a higher statistic and a smaller systematic error has been awaited 
and is necessary, using an independent way from the emulsion study.

Additionally, the $\Lambda$ 
binding energy and low-lying structure of $^{10}_{\Lambda}$Be 
are also relevant to a study of the $\Xi N$ interaction
on which there is almost no experimental information so far.
An event of a bound system of $\Xi^{-}$-$^{14}$N 
decaying into $^{10}_{\Lambda}$Be and $^{5}_{\Lambda}$He
was recently identified 
in an emulsion experiment at KEK~\cite{cite:nakazawa}. 
The analysis of this event was done 
using theoretically predicted energy levels of $^{10}_{\Lambda}$Be 
to obtain the $\Xi^{-}$ binding energy.
This present study provides the energy levels of 
$^{10}_{\Lambda}$Be experimentally for the first time.

\section{Experimental Setup}
\begin{figure}[!htb]
  \begin{center}
    \includegraphics[width=8.6cm]{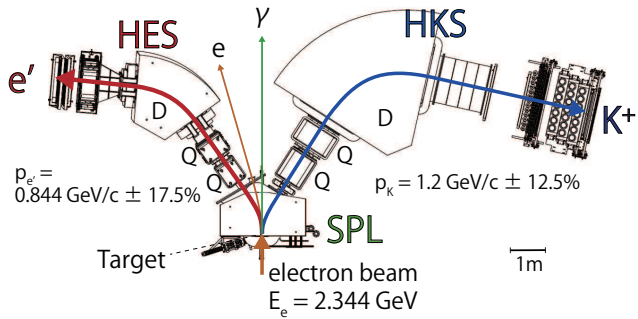}
    \caption{(Color online) A schematic drawing of the JLab E05-115 experimental geometry. 
      The setup consists of SPL, HKS and HES spectrometers. 
      An electron beam with the energy of 2.344~GeV is incident on the 
      target located 
      at the entrance of SPL. A $K^{+}$ and an $e^{\prime}$ 
      with the momenta of $\sim$1~GeV/$c$ are observed 
      by HKS and HES, respectively.}
    \label{fig:setup_e05115}
  \end{center}
\end{figure}

The experiment was performed at JLab Hall C using 
continuous wave electron beams accelerated by CEBAF. 
Fig.~\ref{fig:setup_e05115} shows the experimental 
geometry of JLab E05-115. 
Electrons with the energy of 2.344~GeV were incident on 
a target which was located at the  
entrance of a charge separation dipole magnet (SPL).
Momentum vectors of the reaction $K^{+}$s ($p^{center}_{K}=1.200$~GeV/$c$)
and the scattered electrons ($p^{center}_{e^{\prime}}=0.844$~GeV/$c$) 
at the target were measured by 
the High resolution Kaon Spectrometer (HKS)~\cite{cite:gogami,cite:fujii}
and High resolution Electron spectrometer (HES), respectively. 
The HKS, which has a Q-Q-D magnet configuration, 
was constructed and used in the previous 
{\eek} experiment, JLab E01-011~\cite{cite:7LHe,cite:12LB}. 
The HES, which also has a Q-Q-D magnet configuration, 
and the SPL were newly constructed for the 
JLab E05-115 experiment.
The HKS detector consists of three layers of Time Of Flight (TOF) detectors
for a trigger and off-line Particle IDentification (PID), 
two drift chambers (twelve layers in total) for a particle tracking and 
two types of {\cerenkov} detectors (water ($n=1.33$) and aerogel ($n=1.05$))
for both on-line and off-line PID. 
On the other hand, the HES detector consists of 
two layers of TOF detectors for the trigger and 
two drift chambers (sixteen layers in total) for the particle tracking.
The important feature of the full magnetic spectrometer system 
is the momentum resolution of $\Delta p/p \simeq 2.0 \times 10^{-4}$ 
(FWHM) for both the $e^{\prime}$ and $K^{+}$ spectrometers.
This reaction particle resolution results in
sub-MeV energy resolution (FWHM) of hypernuclear energy levels. 
Refer to \cite{cite:fujii,cite:gogami,cite:12LB} 
for details of the magnets and particle detector systems.


\section{Momentum Reconstruction and Systematic Error on $B_{\Lambda}$}
The momentum vectors of an $e^{\prime}$ and a $K^{+}$ were 
derived by backward transfer matrices which convert
positions and angles at focal plane of each spectrometer to 
momentum vectors at the target. These are then used to obtain the 
residual missing mass of the reaction. 
In our analyses, sixth order backward transfer matrices were used as
they have the minimal complexity to achieve sub-MeV (FWHM) 
energy resolution.
The backward transfer matrices were calibrated~\cite{cite:12LB} with 
the known masses of $\Lambda$ and $\Sigma^{0}$~\cite{cite:pdg} in 
$p${\eek}$\Lambda$,$\Sigma^{0}$ reactions on a 0.45~g/cm$^{2}$ polyethylene target. 
Systematic errors of the binding and 
excitation energies were estimated 
by a full-modeled Monte Carlo simulation which took
into account spectrometer acceptance, particle detector resolutions, 
energy loss, and multiple scattering in all
materials ($e.g.$ target, air, detectors). 
In the simulation, the initial backward transfer matrices were 
perfect, so they were 
distorted in order to produce a more realistic simulation, such as 
peak broadening and shifts in the missing 
mass spectra. The backward transfer matrices were then 
optimized by the same code which was used to optimize the matrices with
real, measured data. 
This procedure was tested and repeated with different sets of 
distorted backward transfer matrices and artificial data. 
As a result, the differences between the assumed and 
obtained values in the binding and excitation energies
were found to be $\leq 0.09$~MeV and 
$\leq0.05$~MeV, respectively. Thus, the
total systematic errors of the binding energy and the excitation 
energy, including target thickness uncertainties were estimated to be
$0.11$~MeV and $0.05$~MeV, respectively.


\section{Results}
\subsection{Definition of the Differential Cross Section}
The hypernuclear electroproduction cross section
can be related to photo-production by virtual photons~\cite{cite:sotona}. 
The virtual photon momentum squared, $Q^{2}=-q^{2}>0$ is 
quite small ($Q^{2} \simeq 0.01$~[(GeV/$c$)$^{2}$], 
transverse polarization $\epsilon_{T} \simeq 0.63$) in our experimental 
geometry. Thus, the ($e,e^{\prime}$) virtual photon can be approximated by a
real photon to obtain the ($\gamma^{*}$,$K^{+}$) differential cross section. This
results in the following equation:
\begin{eqnarray}
  \overline{\Bigl(\frac{d\sigma}{d\Omega_{K}}\Bigr)}
  &=& \frac{\int_{{\rm HKS}}d\Omega_{K} (\frac{d\sigma}{d\Omega_{K}}) }{\int_{{\rm HKS}}d\Omega_{K}} \\
  &=& \frac{1}{N_{T}} \frac{1}{\epsilon^{{\rm HES}} N_{\gamma^{*}}}
  \frac{1}{\epsilon_{{\rm \kappa}}} \sum_{i=1}^{N_{{\rm HYP}}}
  \frac{1}{\epsilon^{{\rm HKS}}_{i}  \Delta \Omega_{i}^{{\rm HKS}}} \label{eq:cross-section}
  \label{eq:cs}
\end{eqnarray}

In the above,  $N_{T}$ is the areal density of target nuclei, 
$N_{\gamma^{*}}$ is the number of virtual photons, 
$N_{{\rm HYP}}$ is the number of $\Lambda$ hypernuclei, 
$\epsilon$'s are correction factors including various efficiencies
(trigger efficiency, detector efficiency, event selection efficiency, 
$K^{+}$ decay factor, $K^{+}$ absorption factor {\it etc.}), and 
$\Delta \Omega_{i}^{{\rm HKS}}$ is the HKS solid angle.
The $\Delta \Omega_{i}^{{\rm HKS}}$ was calculated event by 
event, depending on $K^{+}$ momentum.
We estimated an integral of the virtual photon flux, 
which is defined in \cite{cite:sotona}, 
over the HES acceptance with a Monte Carlo technique.
The integrated virtual photon flux was obtained as $5.67 \times 10^{-5}$~[/electron], 
and it was multiplied by the number of incident electrons 
on the target to evaluate the $N_{\gamma^{*}}$.
The $\Delta\Omega_{i}^{{\rm HKS}}$ and some of correction factors 
were also estimated by Monte Carlo simulations.
It is noted that the differential cross section is 
averaged over the acceptance of our spectrometer system as shown in \cite{cite:12LB}.

\subsection{Missing Mass Spectrum and Peak Fitting}
\begin{figure}[!htb]
  \begin{center}
    \includegraphics[width=8.6cm]{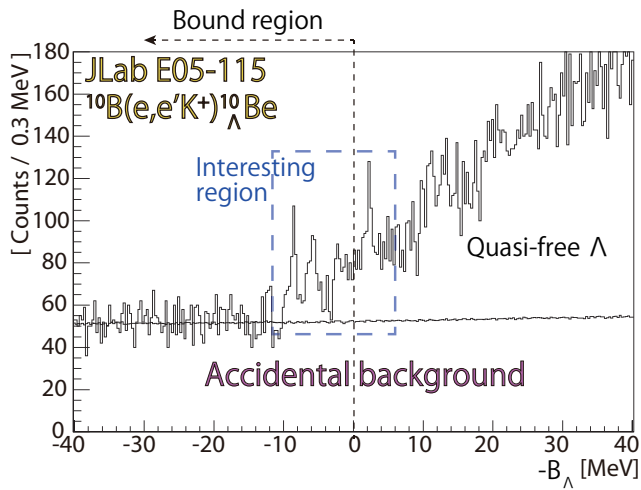}
    \caption{(Color online) The binding energy spectrum of $^{10}_{\Lambda}$Be.
      A distribution of an accidental coincidence between an 
      $e^{\prime}$ and a $K^{+}$
      was obtained by the mixed event analysis as described in the text.}
    \label{fig:LBe10_count}
  \end{center}
\end{figure}

Fig.~\ref{fig:LBe10_count} shows the $\Lambda$ binding energy spectrum 
of $^{10}_{\Lambda}$Be.  The target, a 0.056~g/cm$^{2}$ $^{10}$B foil, was 
isotopically enriched to a purity of 99.9$\%$.
The nuclear masses, 8392.75~MeV/$c^{2}$ and 9324.44~MeV/$c^{2}$, for
$^{9}$Be~\cite{cite:audi} and $^{10}$B~\cite{cite:audi}, respectively, 
were used to calculate the $\Lambda$ binding energy.
The spectrum of accidental coincidences between an $e^{\prime}$ and a $K^{+}$ 
was obtained by the mixed event analysis. This analysis  
reconstructs the missing mass with a random combination 
of $e^{\prime}$ and $K^{+}$ events in each spectrometer acceptance 
in an off-line analysis.
The method gives the accidental-coincidence spectrum 
with higher statistic as much as we need to 
reduce enough an effect of the statistical uncertainty when 
the accidental-coincidence spectrum was subtracted from 
the original missing mass spectrum in the further analysis.

The quasi-free $\Lambda$ ($-B_{\Lambda} \geq 0$)
spectrum was assumed to be represented by a third order polynomial 
function convoluted by a 
Voigt function (convolution of Lorentz and Gauss functions) 
having the experimental energy resolution.
The accidental coincidence and quasi-free $\Lambda$ 
spectra were subtracted from the original binding energy spectrum, and
a test of statistical significance ($=S/\sqrt{S+N}$) was performed to find 
peak candidates. 
\begin{figure*}[!htb]
  \begin{minipage}{0.48\hsize}
    \includegraphics[width=7.5cm]{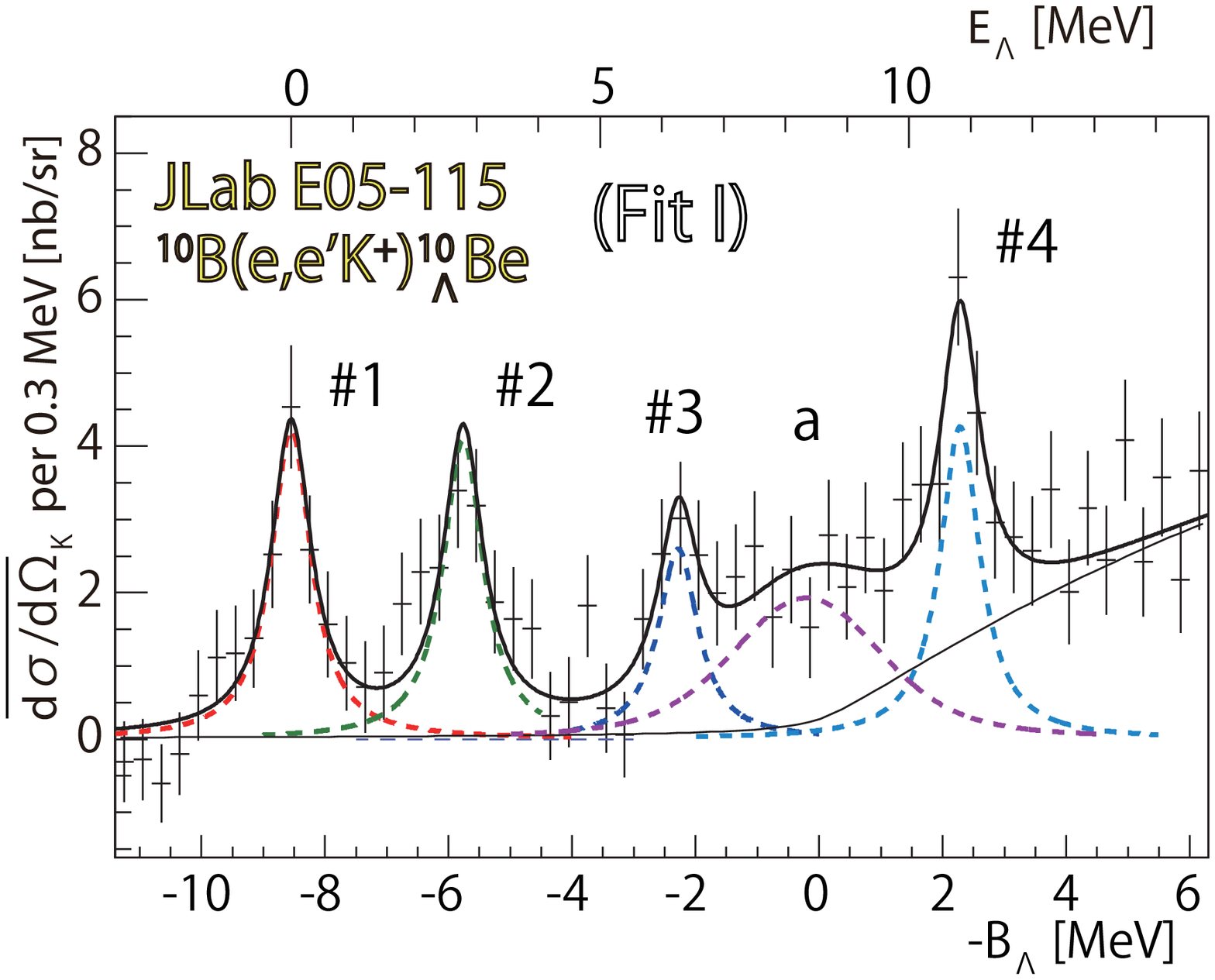}
    \caption{(Color online) The binding energy ($B_{\Lambda}$) and excitation energy 
      ($E_{\Lambda} \equiv -( B_{\Lambda}-B_{\Lambda}(\#1)$) )
      spectra for the $^{10}$B{\eek}$^{10}_{\Lambda}$Be reaction
      with a fitting result of Fit I.
      The ordinate axis is ($\overline{d\sigma/d\Omega_{K}}$) per 0.3~MeV.}
    \label{fig:LBe10csfit}
  \end{minipage}\hspace{0.2cm}
  \begin{minipage}{0.48\hsize}
    \includegraphics[width=7.5cm]{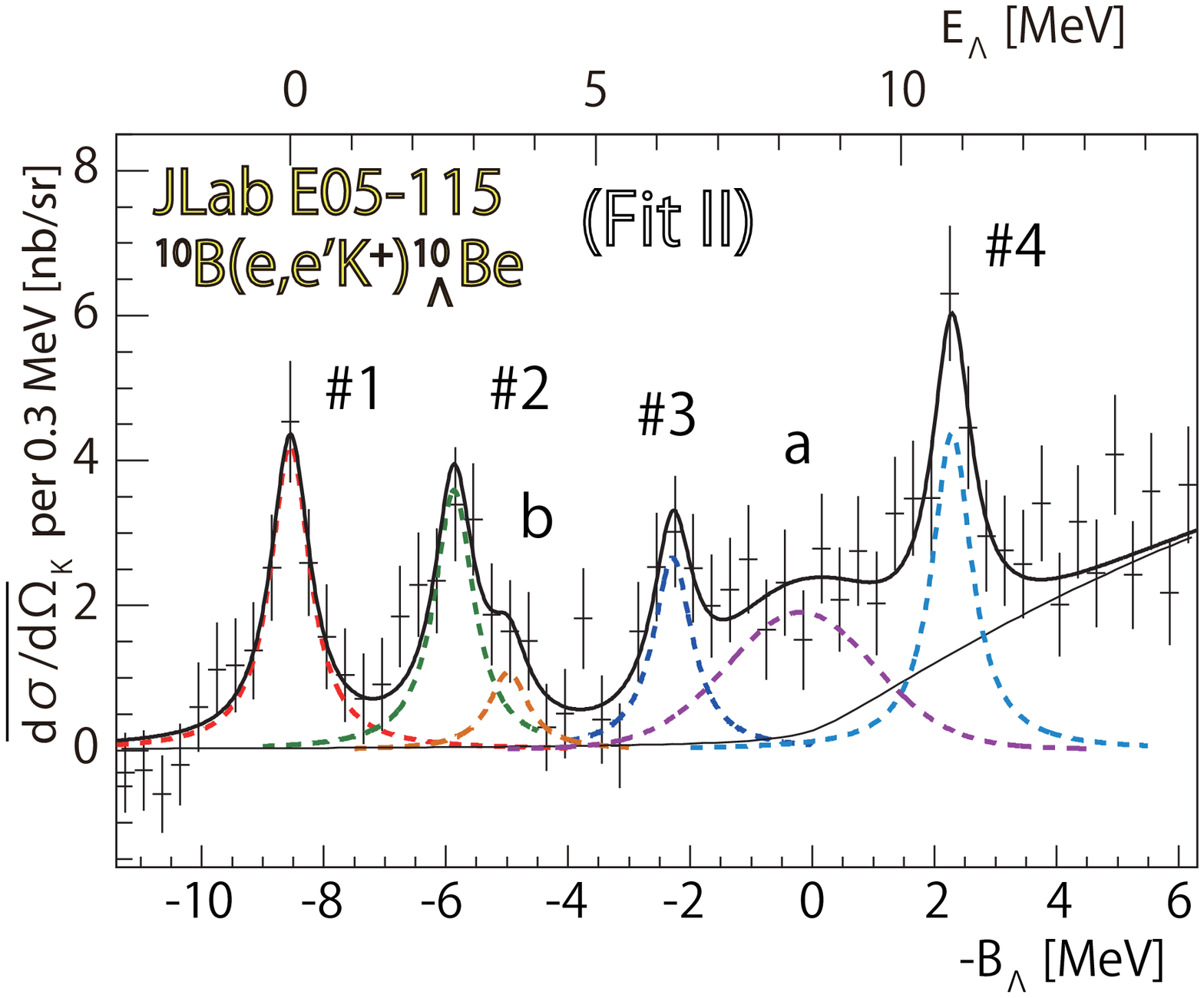}
    \caption{(Color online) The binding energy ($B_{\Lambda}$) and excitation energy 
      ($E_{\Lambda} \equiv -( B_{\Lambda}-B_{\Lambda}(\#1) )$ )
      spectra for the $^{10}$B{\eek}$^{10}_{\Lambda}$Be reaction
      with a fitting result of Fit I\hspace{-.1em}I.
      The ordinate axis is ($\overline{d\sigma/d\Omega_{K}}$) per 0.3~MeV.}
    \label{fig:LBe10csfit2}
  \end{minipage}
\end{figure*}
Fig.~\ref{fig:LBe10csfit} shows the binding energy spectrum with 
the ordinate axis of 
($\overline{d\sigma/d\Omega_{K}}$), 
as defined by Eq.~(\ref{eq:cs}).
A fitting result 
by Voigt functions for peak candidates with 
statistical significance of $\geq 5 \sigma$
are also shown in the figure.
The peak candidates are labeled as $\# 1$, $\# 2$, $\# 3$ and $\# 4$,
and are identified as candidates of hypernuclear states.
\begin{table*}[!htb]
  \begin{center}
    \caption{The binding energies ($B_{\Lambda}$), 
      excitation energies ($E_{\Lambda} \equiv -( B_{\Lambda}-B_{\Lambda}(\#1)$) ) 
      and differential cross sections for 
      $^{10}$B($\gamma^{*},K^{+}$)$^{10}_{\Lambda}$Be.
      The error is statistical. 
      The systematic errors on $B_{\Lambda}$ and $E_{\Lambda}$ are
      0.11~MeV and 0.05~MeV, respectively.
      The systematic errors on the differential cross sections are $\pm9\%$
      for $\#1,2,3,b$, (${+10\%}/{-9\%}$) for $\#4$, and (${+42\%}/{-9\%}$) for $a$.
    }
    \label{tab:LBe10_results}
    \begin{tabular}{|c|c|c|c||c|c|c|}
      \hline \hline
      ID& \multicolumn{3}{c||}{Fit I} & \multicolumn{3}{c|}{Fit I\hspace{-.1em}I} \\ \cline{2-7}
      &  $-B_{\Lambda}$~[MeV] & $E_{\Lambda}$~[MeV]
      & $\overline{\Bigl(\frac{d\sigma}{d\Omega_{K}}\Bigr)}$~[nb/sr]
      & $-B_{\Lambda}$~[MeV] & $E_{\Lambda}$~[MeV]
      & $\overline{\Bigl(\frac{d\sigma}{d\Omega_{K}}\Bigr)}$~[nb/sr]\\ \hline \hline
      
      $\#$1 &  -8.55$\pm$0.07 & 0.0 & 17.0$\pm$0.5
            &  -8.55$\pm$0.07 & 0.0 & 17.1$\pm$0.5\\ \hline 
      $\#$2 &  -5.76$\pm$0.09 & 2.78$\pm$0.11 & 16.5$\pm$0.5
            &  -5.87$\pm$0.18& 2.68$\pm$0.19 & 14.5$\pm$0.4\\ \hline 
      $\#$3 & -2.28$\pm$0.14 & 6.26$\pm$0.16 & 10.5$\pm$0.3
            & -2.29$\pm$0.14 & 6.26$\pm$0.15 & 10.7$\pm$0.3\\ \hline
      $\#$4 & +2.28$\pm$0.07 & 10.83$\pm$0.10 & 17.2$\pm$0.5
            & +2.29$\pm$0.07 & 10.83$\pm$0.10& 17.7$\pm$0.5 \\ \hline 
      $a$   & -0.20$\pm$0.40 &  8.34$\pm$0.41& 23.2$\pm$0.7
            & -0.19$\pm$0.38 & 8.36$\pm$0.39 & 20.5$\pm$0.6\\ \hline 
      $b$   & & & & -4.98$\pm$0.53 & 3.57$\pm$0.53 & 4.4$\pm$0.1\\ 
            \hline \hline
    \end{tabular}
  \end{center}
\end{table*}

\begin{figure}[!htb]
  \begin{center}
    \includegraphics[width=8.6cm]{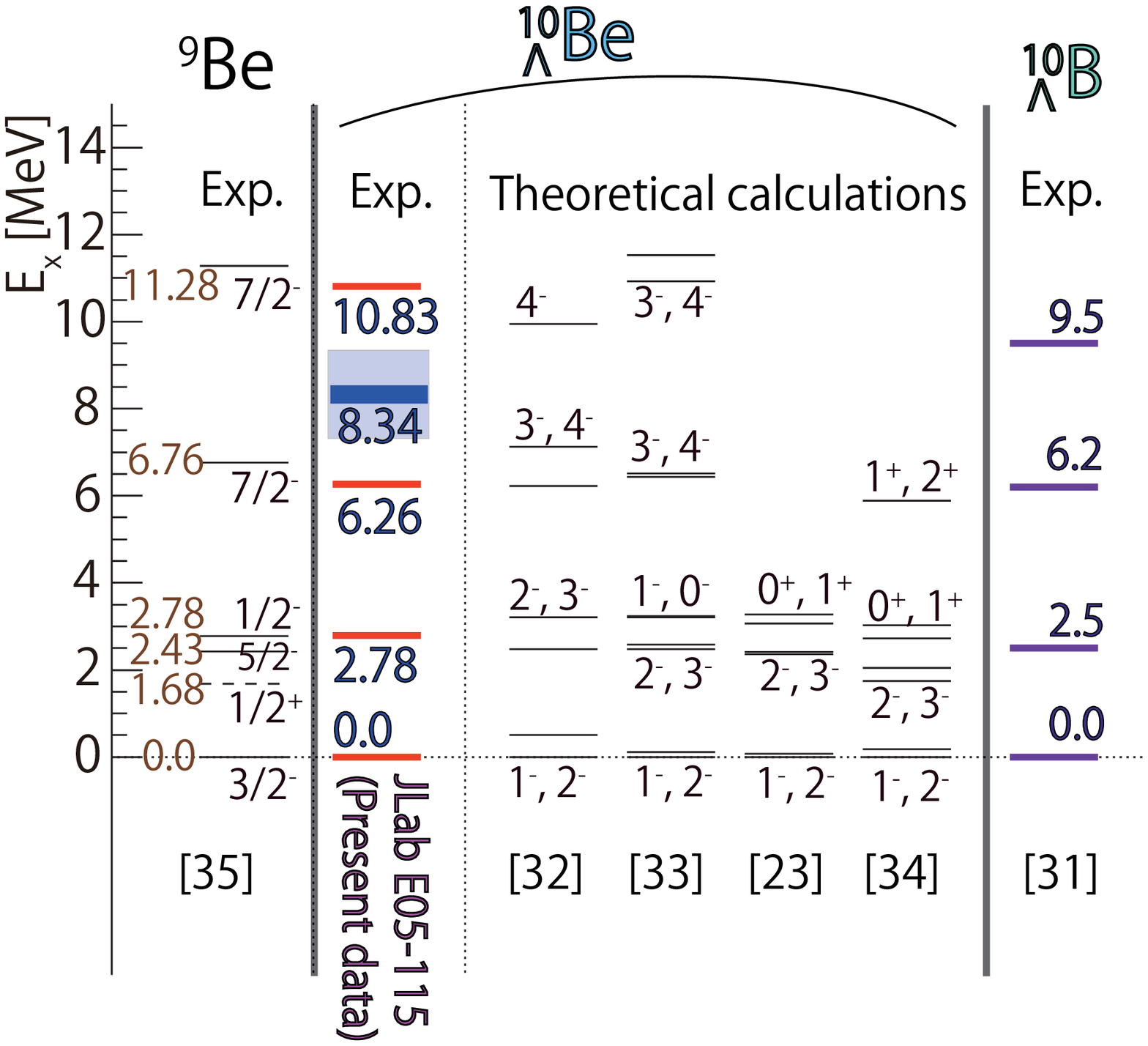}
    \caption{(Color online) The obtained energy levels of $^{10}_{\Lambda}$Be (Fit I)
      compared to those of the theoretical 
      calculations~\cite{cite:hiyama_10LB,cite:motoba2,cite:millener,cite:isaka_10LBe}.
      The energy levels of $^{9}$Be~\cite{cite:tilley} and $^{10}_{\Lambda}$B~\cite{cite:hasegawa2}
      are shown for comparisons as well.}
    \label{fig:LBe10_comp}
  \end{center}
\end{figure}

The enhancements between the peaks of $\# 3$ and $\# 4$ 
are considered to be several states and 
were included by fitting with a shape having broader width 
(indicated as $a$ in Fig.~\ref{fig:LBe10csfit}). The
FWHMs of the Voigt functions for the peaks of $\# 1$-$4$ and $a$
were found to be 0.78~MeV and 2.87~MeV, respectively.
The 0.78 MeV (FWHM) resolution is almost three times 
better than the measurement of its mirror $\Lambda$ 
hypernucleus, $^{10}_{\Lambda}$B  measured at KEK
(2.2 MeV FWHM) using the {\pik} reaction~\cite{cite:hasegawa2}.
The fitted results are summarized in Table.~\ref{tab:LBe10_results} as Fit I.
The statistical error is given in the results.
Fig.~\ref{fig:LBe10_comp} shows the measured  
excitation energy levels (Fit I), 
the theoretical calculations of 
$^{10}_{\Lambda}$Be~\cite{cite:hiyama_10LB,cite:motoba2,cite:millener,cite:isaka_10LBe},  
and the experimental results of $^{9}$Be~\cite{cite:tilley} 
and $^{10}_{\Lambda}$B~\cite{cite:hasegawa2}.
The differential cross section of each state 
for the $^{10}$B($\gamma^{*}$,$K^{+}$)$^{10}_{\Lambda}$Be 
reaction relates to that of a spectroscopic factor ($C^{2}S$) of the proton pickup reaction 
from $^{10}$B. The $C^{2}S$ of $^{10}$B($e,e^{\prime}$p)$^{9}$Be are
reported in \cite{cite:tilley}, and 
they are $1.000, 0.985, 0.668$ and $ 1.299$ for 
$J^{\pi}=3/2^{-}$, $5/2^{-}$, $7/2_{1}^{-}$ and $7/2_{2}^{-}$ states in $^{9}$Be, respectively.
Comparing energy levels and the differential cross sections 
of hypernuclear states (Table.~\ref{tab:LBe10_results})
with energy levels of $^{9}$Be (Fig.~\ref{fig:LBe10_comp}) 
and $C^{2}S$ of $^{10}$B($e,e^{\prime}$p)$^{9}$Be, 
the peaks of $\#1$, $\#2$ $\#3$ and $\#4$ 
respectively correspond to $J^{\pi}=3/2^{-}$, $5/2^{-}$ 
$7/2_{1}^{-}$ and $7/2_{2}^{-}$ states in $^{9}$Be.
In the theoretical predictions 
of $^{10}_{\Lambda}$Be energy levels
shown in Fig.~\ref{fig:LBe10_comp},
the states of $0^{-}$, $1^{-}$ 
($^{9}$Be$[J^{\pi};E_{x}]\otimes j^{\Lambda} = [$1/2$^{-}$; 2.78~MeV$]\otimes s^{\Lambda}_{1/2}$)
and $0^{+}$, $1^{+}$ 
($^{9}$Be$[J^{\pi};E_{x}]\otimes j^{\Lambda} = [$1/2$^{+}$; 1.68~MeV$]\otimes s^{\Lambda}_{1/2}$)
are predicted to be above the $2^{-}$, $3^{-}$ states
($^{9}$Be$[J^{\pi};E_{x}]\otimes j^{\Lambda} = [$5/2$^{-}$; 2.43~MeV$]\otimes s^{\Lambda}_{1/2}$) 
by about 1~MeV. 
There might be a possibility that these states 
are at around 1~MeV above the peak of $\#2$, assuming the peak of $\#2$ corresponds to 
the $2^{-}$ and $3^{-}$ states.
Thus, a fitting with an additional peak function (labeled as $b$) 
with a width of 0.78~MeV (FWHM) around 1~MeV above 
the peak of $\#2$ was also performed,
and the fitting result is shown in Table.~\ref{tab:LBe10_results}
(labeled as Fit I\hspace{-.1em}I) and Fig.~\ref{fig:LBe10csfit2}.

A systematic error on the cross section come from 
uncertainties of trigger efficiency, analysis efficiencies 
such as tracking and event selection,
correction factors such as the solid angle of 
spectrometer system and $K^{+}$ decay factor, 
and so on. 
A square-root of sum of squares of these uncertainties 
was obtained to be $9\%$, and it is used as 
the systematic error on the differential cross section. 
Obtained differential cross sections of the peak $\#4$ and $a$ 
depend on the assumption of quasi-free $\Lambda$ distribution in the fitting.
We tested usages of lower-order polynomial functions (first and second orders)
for the quasi-free $\Lambda$ events in order to estimate additional systematic errors
for the peak $\#4$ and $a$.
As results, the differential cross sections for peak $\#4$ and $a$ 
were changed by $\leq+5\%$ and $\leq+41\%$, respectively, 
although the others were not changed within the statistical errors.
Therefore, the systematic errors on the differential cross sections for 
the peak $\#4$ and $a$ were estimated to be (${+10\%}/{-9\%}$) and 
(${+42\%}/{-9\%}$), respectively.
It is noted that, in the test, obtained peak-means for all of peak candidates 
did not vary within the statistical errors. 

\section{Discussion}
The peak of $\# 1$ is
identified as the sum of the 1$^{-}$ and 2$^{-}$ states
($^{9}$Be$[J^{\pi};E_{x}]\otimes j^{\Lambda} = [$3/2$^{-}$; g.s.$]\otimes s^{\Lambda}_{1/2}$). 
The energy spacing between these two states is expected to be less than $\sim 0.1$~MeV 
according to a gamma-ray measurement of 
the mirror hypernucleus, $^{10}_{\Lambda}$B~\cite{cite:tamura_bnl} and theoretical 
calculations~\cite{cite:hiyama_10LB,cite:millener}. 
The differential cross sections for these states were 
predicted to be comparable in DWIA calculation~\cite{cite:motoba3}.
We performed a simulation varying 
the cross-section ratio of $1^{-}$ to that of $2^{-}$
from 0.5 to 1.5, and also the energy separation 
between these two states from 0.05 to 0.15~MeV,
in order to estimate the ground state binding energy.
As a result of the simulation, 
the ground state binding energy would be: 
\begin{eqnarray}
  B_{\Lambda}^{g.s.} (^{10}_{\Lambda}{\rm Be}) 
  &=& B_{\Lambda} (\#1) + C_{1},
\end{eqnarray}
where $C_{1}=0.05\pm0.05$~MeV.
Thus,
\begin{eqnarray}
  B_{\Lambda}^{g.s.} (^{10}_{\Lambda}{\rm Be}) 
  &=& (8.55+0.05) \pm 0.07^{stat.}  \pm (0.11+0.05)^{sys.}\nonumber \\
  \\
  &=& 8.60 \pm 0.07^{stat.}  \pm 0.16^{sys.}~{\rm MeV}.
\end{eqnarray}


\begin{figure}[!htbp]
  \begin{center}
    \includegraphics[width=8.6cm]{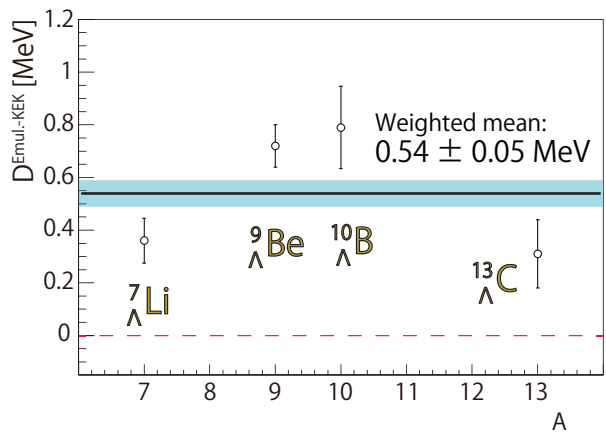}
    \caption{(Color online) The binding energy differences 
      of $^{7}_{\Lambda}$Li, $^{9}_{\Lambda}$Be, $^{10}_{\Lambda}$B and $^{13}_{\Lambda}$C
      between the emulsion experiments~\cite{cite:12LC_2} 
      and the {\pik} experiments~\cite{cite:hashimototamura} with the statistical errors. 
      The values of {\pik} experiments were subtracted from those of the emulsion experiments.
      The weighted mean was obtained to be $+0.54 \pm 0.05$~MeV as 
      represented by a solid line.
      The plots should be on a line of zero (dashed line) if the binding energies 
      measured in the {\pik} and emulsion experiments are consistent.     
    }
    \label{fig:emul-kek}
  \end{center}
\end{figure}
A comparison of the binding energy of $^{10}_{\Lambda}$Be 
with its isotopic mirror hypernucleus, $^{10}_{\Lambda}$B 
provides an information of the $\Lambda N$ CSB effect. 
Before comparing the current results with previous measurements, 
we note that there appears to be a systematic discrepancy 
between the binding energies obtained from emulsion experiments and 
those obtained from {\pik} experiments.
This is illustrated in Fig.~\ref{fig:emul-kek}, which 
shows the binding energy differences 
between the emulsion experiments~\cite{cite:12LC_2} 
and the {\pik} experiments~\cite{cite:hashimototamura}.
The binding energy of $^{10}_{\Lambda}$B was measured by 
both the emulsion experiment~\cite{cite:12LC_2} 
and the {\pik} experiment~\cite{cite:hasegawa2}.
The reported binding energies of $^{10}_{\Lambda}$B 
obtained by the emulsion experiment and {\pik} experiment 
are respectively $8.89\pm0.06^{stat.}\pm0.04^{sys.}$~MeV 
and $8.1\pm0.1^{stat.}\pm0.5^{sys.}$~MeV, 
which are not consistent.
In the {\pik} experiment, however, the binding energy was derived
using a reference of binding energy of $^{12}_{\Lambda}$C which 
was measured by the emulsion experiments~\cite{cite:12LC_1,cite:12LC_2}. 
There are binding energy data of 
$^{7}_{\Lambda}$Li, $^{9}_{\Lambda}$Be, $^{10}_{\Lambda}$B and $^{13}_{\Lambda}$C
by the {\pik} experiments using the $^{12}_{\Lambda}$C reference 
to be compared with those measured in the emulsion experiments.
The binding energy differences 
of $^{7}_{\Lambda}$Li, $^{9}_{\Lambda}$Be, $^{10}_{\Lambda}$B and $^{13}_{\Lambda}$C
between the emulsion experiments~\cite{cite:12LC_2} 
and the {\pik} experiments~\cite{cite:hashimototamura} with the statistical errors 
are respectively $D_{{\rm Emul.-KEK}}=+0.36 \pm 0.09$, $+0.72 \pm 0.08$,
$+0.79 \pm0.16$ and $+0.31 \pm 0.13$~MeV as shown in Fig.~\ref{fig:emul-kek}.
It seems there is a systematic energy difference, 
and the difference
is found to be $D_{{\rm Emul.-KEK}}^{fit}=+0.54\pm0.05$~MeV 
by the weighted mean of these four points as represented by a
solid line in Fig.~\ref{fig:emul-kek}.
It indicates that the reported binding energy of $^{12}_{\Lambda}$C is 
shifted by about $(C_{2}\equiv)$~0.54~MeV.
If the binding energy of $^{10}_{\Lambda}$B measured 
by the {\pik} experiment is corrected with $C_{2}$, 
it becomes consistent with that of emulsion experiment within the errors.
Since the results from {\pik} experiments are all calibrated to 
the earlier $^{12}_{\Lambda}$C binding energy measurement, 
it is prudent to consider the possibility that they should be 
recalibrated to the emulsion results for the four hypernuclei shown 
in Fig.~\ref{fig:emul-kek}.

\begin{table*}[htbp]
  \begin{center}
    \caption{Corrected binding energies to be compared with each other for $A=10$
      and $A=12$ $\Lambda$ hypernuclei. The errors are statistical. 
      The systematic errors are 0.16~MeV for $^{10}_{\Lambda}$Be (present data), 
      0.11~MeV for $^{12}_{\Lambda}$B (JLab),  
      0.04~MeV for $^{10}_{\Lambda}$Be, $^{10}_{\Lambda}$B, $^{12}_{\Lambda}$C and $^{12}_{\Lambda}$B measured
      by the emulsion experiments, 
      0.50~MeV for $^{10}_{\Lambda}$B measured 
      by the {\pik} experiment at KEK.}
    \label{tab:bl_diff10-12}
    \begin{tabular}{|c|c|c|c|c|}
      \hline \hline
      Hypernucleus &Experiment& Reported $B^{g.s.}_{\Lambda}$~[MeV] 
      & Correction~[MeV] & Corrected $B^{g.s.}_{\Lambda}$~[MeV] \\ \hline \hline
      $^{10}_{\Lambda}$Be & Present data & $8.60\pm0.07$ & - & $8.60\pm0.07$ \\ 
      & Emulsion~\cite{cite:juric,cite:cantwell} & $9.11\pm0.22$ & - & $9.11\pm0.22$ \\ \hline
      $^{10}_{\Lambda}$B & KEK~\cite{cite:hasegawa2} & $8.1\pm0.1$ & $C_{2}=+0.54$ & $8.64\pm0.1$ \\  
      & Emulsion~\cite{cite:12LC_2} & $8.89\pm0.12$ & - & $8.89\pm0.12$ \\  \hline \hline
      $^{12}_{\Lambda}$B & JLab~\cite{cite:12LB} & $11.529\pm0.025$ & - & $11.529\pm0.025$ \\  
                         & Emulsion~\cite{cite:12LC_2} & $11.37\pm0.06$ & - & $11.37\pm0.06$ \\   \hline
      $^{12}_{\Lambda}$C & Emulsion~\cite{cite:12LC_1,cite:12LC_2} & $10.76\pm0.19$ & $C_{2}$  & $11.30\pm0.19$ \\ \hline \hline
    \end{tabular}
  \end{center}
\end{table*}
The ground state binding energies of $^{10}_{\Lambda}$B and 
$^{10}_{\Lambda}$Be to be compared from each other,
taking into account the above 
corrections ($C_{1,2}$) are summarized in Table.~\ref{tab:bl_diff10-12}.
The present result shows that the ground state binding energy 
of $^{10}_{\Lambda}{\rm Be}$ is shallower than 
the weighted-mean value of three events reported in the emulsion experiments 
by $0.51\pm0.23$~MeV, and differences of the ground state binding energies 
between $^{10}_{\Lambda}{\rm B}$ and $^{10}_{\Lambda}{\rm Be}$ were 
obtained to be $\Delta B_{\Lambda}(^{10}_{\Lambda}{\rm B}-^{10}_{\Lambda}{\rm Be})
=0.04\pm0.12$~MeV (KEK after the $C_{2}$ correction and JLab) 
and $0.29\pm0.14$~MeV (Emulsion and JLab).
The obtained binding energy would be a considerable constraint 
for the study of $\Lambda N$ CSB interaction 
in the $A=10$, $T=1/2$ iso-doublet $\Lambda$ hypernuclear system 
since the effect on the binding energy difference among 
mirror hypernuclei is expected to be a few hundred keV level 
or less~\cite{cite:hiyama_10LB,cite:gal}. 
For example, the effect of $\Lambda N$ CSB interaction 
on the binding energy difference between $^{10}_{\Lambda}$Be and $^{10}_{\Lambda}$B is 
predicted to be only $0.2$~MeV by the four-body cluster model with 
the phenomenological even-state $\Lambda N$ CSB potential~\cite{cite:hiyama_10LB}.

In the above discussion, the correction of $C_{2}$ ($=0.54$~MeV) was used for 
the {\pik} result. 
The $0.54$~MeV shift of the reported binding energy of 
$^{12}_{\Lambda}$C gives a great impact since 
it was used as a reference of the binding energy measurements 
for all the {\pik} experiments in which most energy levels of 
$\Lambda$ hypernuclei with $A>16$ were obtained and used 
as theoretical inputs for the study of $\Lambda N$ potential. 
Therefore, well-calibrated binding energy measurements particularly for 
medium to heavy mass region are needed, and 
only the {\eek} experiment would be suitable for the purpose 
at the moment. 

Moreover, the 0.54~MeV shift
changes a situation of the binding energy difference in $A=12$ 
isotopic mirror $\Lambda$ hypernuclear system.  
Previous published results indicate that there is 
a large binding energy difference between 
$^{12}_{\Lambda}$C and $^{12}_{\Lambda}$B as shown in Table.~\ref{tab:bl_diff10-12}. 
The large difference of $>0.6$~MeV is hard to be explained 
theoretically, and is considered to be 
caused by unexpectedly large $\Lambda N$ CSB effect.
However, the binding energy difference 
becomes $\Delta B_{\Lambda}$ ($^{12}_{\Lambda}$C $-^{12}_{\Lambda}$B) $ = -0.23 \pm 0.19$~MeV 
(Emulsion and JLab) if the correction of $C_{2}=0.54$~MeV is applied. 
It shows that the binding energy difference between isotopic 
mirror hypernuclei in $A=12$ system is 
less than a few hundred keV as well as that in $A=10$ system, 
implying the effect of $\Lambda N$ CSB would be small 
in $p$-shell $\Lambda$ hypernuclei as expected in 
the theoretical prediction~\cite{cite:gal}.

\section{Summary}
This paper reports on a high resolution {\eek} experiment which obtained for
the first time the energy spectrum of the hypernucleus, $^{10}_{\Lambda}$Be.
The experiment which used a new magnetic spectrometer system 
was successfully carried out to obtain 
hypernuclear structures with an energy resolution of $\sim 0.78$~MeV (FWHM) and
a systematic error of 0.11~MeV on the binding energy measurement. 
The $\Lambda$ binding energy of the first doublet ($1^{-}$, $2^{-}$) was 
obtained to be $8.55 \pm 0.07$~MeV.
The result implies that the ground state is 
shallower than that reported in the emulsion experiment by about 0.5~MeV,
 which would provide insights for 
the study of the CSB effect in the $\Lambda N$ interaction.
In the discussion of the binding energy difference 
between $A=10$, $T=1/2$ iso-doublet hypernuclei, 
a correction of $0.54$~MeV on $^{12}_{\Lambda}$C which 
is indicated by the emulsion and {\pik} experiments was used.
The 0.54~MeV correction on $^{12}_{\Lambda}$C 
binding energy makes results in the emulsion and {\pik} experiments 
consistent, and support a small $\Lambda N$ CSB effect 
in the $A=10$ and $A=12$ hypernuclear systems.

The present result demonstrated that 
the {\eek} experiment is able to measure
energy levels with a better accuracy 
in addition to higher resolution than 
that of the hadron missing mass spectroscopy, 
and it would be a powerful tool to 
investigate energy levels of $\Lambda$ hypernuclei
particularly for medium to heavy mass region in the future.

\section*{Acknowledgments}
We thank the JLab staffs of the physics, accelerator, and
engineer divisions for support for the experiment. 
Also, we thank E.~Hiyama, M.~Isaka, D.J.~Millener, Y.~Yamamoto,
and T.~Motoba for intensive discussions related to their theoretical works. 
The program at JLab Hall-C is supported by JSPS KAKENHI Grant 
No.~12002001, No.~15684005, No.~16GS0201, and No.~244123 (Gant-in-Aid for JSPS fellow),
JSPS Core-to-Core Program No.~21002, and 
JSPS Strategic Young Researcher Overseas Visits
Program for Accelerating Brain Circulation No.~R2201.
This work is supported by U.S. Department of Energy contracts 
No.~DE-AC05-84ER40150, No.~DE-AC05-06OR23177, No.~DE-FG02-99ER41065, 
No.~DE-FG02-97ER41047, No.~DE-AC02-06CH11357, No.~DE-FG02-00ER41110, and 
No.~DE-AC02-98CH10886, and US-NSF contracts No.~013815 and No.~0758095.

\end{document}